\documentclass[aps,prd,preprint,showpacs,tightenlines,eqsecnum,nobibnotes,
nofootinbib]{revtex4}
\usepackage{latexsym,amsmath2000,hyperref}

\newcommand{\be}{\begin{equation}}
\newcommand{\ee}{\end{equation}}
\newcommand{\bea}{\begin{eqnarray}}
\newcommand{\eea}{\end{eqnarray}}

\begin{document}

\title{Entropy Bounds in $R\times S^3$ Geometries}


\author{Iver Brevik}
\email{iver.h.brevik@mtf.ntnu.no}
\affiliation{Division of Applied Mechanics, Norwegian University of Science
and Technology, N-7491 Trondheim, Norway}

\author{Kimball A. Milton}
\email{milton@mail.nhn.ou.edu}
\homepage{www.nhn.ou.edu/
\affiliation{Department of Physics and Astronomy, The University of 
Oklahoma, Norman 73019 USA}

\author{Sergei D.  Odintsov}
\email{odintsov@mail.tomsknet.ru} 
\affiliation{Tomsk State Pedagogical University, 
634041 Tomsk, Russia}

\date{\today}

\begin{abstract}
Exact calculations are given for the Casimir energy for various fields
in $R\times S^3$ geometry.  The Green's function method naturally gives
a result in a form convenient in the high-temperature limit, while the
statistical mechanical approach gives a form convenient for low temperatures.
The equivalence of these two representations is demonstrated.  Some
discrepancies with previous work are noted.  In no case, even for ${\cal N}=4$
SUSY, is the ratio of entropy to energy found to be bounded.
This deviation, however, occurs for low temperature, where the equilibrium
approach may not be relevant. The same methods are used to calculate the energy
and free energy for the TE modes in a half-Einstein universe bounded by a
perfectly conducting 2-sphere.
\end{abstract}

\preprint{OKHEP-02-01}

\maketitle

\section{Introduction}

The remarkable appearance of the holographic principle has fostered 
the understanding that some hitherto distant branches of
theoretical physics may have a much deeper common origin 
than was expected. One outstanding example of this sort is the relation,
suggested by Verlinde \cite{verlinde}, 
between the Cardy entropy formula \cite{cardy} and
the Friedmann equation for the 
evolution of the scale factor of the universe.
Moreover, the proposal that there exists a
holographic bound on the cosmological entropy associated with Casimir energy
(for a general introduction to that subject, see 
Refs.~\cite{mono, elizalde}) suggests
that there should be a deeper relation between Friedmann cosmology and
the Casimir effect. 

There has much activity in the study of the Cardy-Verlinde formula 
and its various applications (see, for example, Refs.~\cite{kl,
odintsov,klemm, cosmology,others} and references therein).
Specifically, there has been much interest in studying the entropy and energy
arising from quantum and thermal fluctuations in conformal field theories
\cite{verlinde,kl,klemm}.  As we remarked, particularly interesting
 is the Verlinde bound \cite{verlinde}
for the ratio of the entropy to the thermal energy.  Whether this bound
can be realized in realistic situations is a matter for specific calculations.
Previous computations \cite{kl,klemm} have been limited to the specific
regime of high temperatures, so they are unable to provide definitive
results.  Here we obtain exact results for various fields in the $R\times S^3$
geometry, so the issues may be more decisively addressed.
We also will consider electromagnetic modes in a half-Einstein universe,
which give similar results using the same methods.

The outline of this paper is as follows.  In the following section we 
consider a conformally coupled scalar in the $R\times S^3$ geometry.
We obtain exact results for the Casimir energy and free energy using
both Green's function and statistical-mechanical methods.  The former
result is most convenient for the high-temperature regime, while the
latter is suited to low temperatures.  However, the results
obtained by the two methods are
exactly equal, as may be shown either by the use the the Euler-Maclaurin
sum formula, or by the Poisson sum formula. For the physical applications
to the evolution of the universe, except, perhaps, for the very earliest
stages, the high-temperature results are the
most relevant.  In Sec.~\ref{sec:d2scalar}, 
for comparison, we compute the same results
for $R\times S^1$.  Vectors and fermions are treated in Secs.~\ref{sec:IV} 
and \ref{sec:V},
respectively.  Entropy bounds are the subject of Sec.~\ref{sec:VI}.  
In the presumably
physically relevant high-temperature regime, the ratio of entropy to energy
is small, of order $(aT)^{-1}$, where $T$ is the temperature and $a$ is the
radius of the sphere.  Linear temperature terms in the energy, omitted
in earlier analyses \cite{klemm}, preclude the existence of an entropy
bound even in the high-temperature regime.
At low temperature, moreover, both the entropy
and the thermal
deviation of the energy from the zero-point energy become exponentially
small, while the ratio of these two quantities diverges as $T\to0$.
In Sec.~\ref{sec:VII} we consider the half-Einstein universe.
Some summarizing remarks are offered in the Conclusions.
There are three Appendices.  The first points out the fundamental difficulty
of breaking conformal symmetry in such problems, the second shows
the applicability of the Euler-Maclaurin method to the example treated in
Sec.~\ref{sec:VII}, and the third responds to a suggestion that in $R\times S^1$
the zero-mode contributes a linear temperature term at low temperature.

\section{Conformally Coupled Scalar}
\label{sec:II}
\subsection{Green's Function Method}
\label{klgfsec}
We can start from the formalism given in Kantowski and Milton \cite{km},
which in turn was based on the Green's function formulation first
given by Schwinger et al.~\cite{js}.
The energy is given by the imaginary part of the Green's function,
\be
U=V_3\partial^0\partial^{\prime0}\Im G(x,y;x',y')\big|_{x=x',y=y'},
\ee
where the ``external'' coordinates $x$ consist only of the time.  We introduce a
Fourier transform there
\be
G(t,y;t',y')=\int\frac{d\omega}{2\pi} e^{-i\omega(t-t')}g(y,y';\omega),
\ee
so
\be
U=-\frac{iV_3}{4\pi}\int_c d\omega\,\omega^2 g(y,y;\omega),
\ee
where the contour $c$ encircles the poles on the positive axis in a
negative sense, and those on the negative axis in a positive sense.
The reduced Green's function satisfies
\be
(\nabla_3^2+\omega^2)g(y,y';\omega)=-\delta(y-y').
\label{rgfeqn}
\ee
The eigenvectors and eigenvalues of the Laplacian on $S^3$ are
\be
\nabla_3^2 Y_l^m(y)=-\frac{M_l^2}{a^2}Y_l^m,\quad M_l^2=l(l+2),
\ee
which have degeneracy $D_l=(l+1)^2$.  The addition theorem is
\be
\sum_mY_l^m(y)Y_l^{m*}(y)=\frac{D_l}{V_3},
\ee
so in view of the eigenfunction construction
\be
g(y,y';\omega)=\sum_{lm}\frac{Y_l^m(y)Y_l^{m*}(y')}{M_l^2/a^2-\omega^2}
\ee
the Casimir energy is
\be
U=-\frac{i}{4\pi}\int_c d\omega\,\omega^2\sum_l\frac{D_l}{M_l^2/a^2-\omega^2}.
\label{ce}
\ee

Conformal coupling in incorporated by replacing the Laplacian in 
Eq.~(\ref{rgfeqn}) by $\nabla_3^2+\xi R$, which for the conformal value
$\xi=1/6$ amounts to the
addition of 1 to the $M_l^2$ operator,
that is, replacing $M_l^2\to(l+1)^2$.

Temperature dependence is incorporated by the replacement
\be
\int_c \frac{d\omega}{2\pi}\to \frac{4i}{\beta}\sum_{n=0}^\infty{}',\quad
\omega^2\to -\left(\frac{2\pi n}{\beta}\right)^2.
\label{finitet}
\ee
The prime on the summation sign means that the $n=0$ term is counted with half
weight.

We carry out the sum on $l$ in Eq.~(\ref{ce}) as follows:
The general representation
\be
\sum_{m=0}^\infty \frac1{m^2-\alpha^2}=-\frac\pi{2\alpha}\cot\pi\alpha
-\frac1{2\alpha^2}
\label{cotan}
\ee
becomes here
\be
\sum_{l=0}^\infty\frac{(l+1)^2}{(l+1)^2/a^2-\omega^2}=\mbox{constant}
+\frac\pi2 i\omega a^3
+\frac{\pi\omega a^3 i}{e^{-2\pi i\omega a}-1}.
\ee
We then make the finite-temperature replacements (\ref{finitet}) and obtain,
after dropping the contact term arising from the constant ($\propto\zeta(-2)
=0$) 
\be
U=\frac1{a}\left(\frac{2\pi a}{\beta}\right)^4
\left[\sum_{n=0}^\infty{}'\frac{n^3}{e^{4\pi^2a n/\beta}
-1}+\frac1{240}\right],
\label{klft}
\ee
which, since the summand vanishes at $n=0$, gives only exponentially small
corrections to Stefan's law,
\be
U\sim\frac1a\frac{(2\pi aT)^4}{240},\quad aT\gg1.
\label{sl}
\ee
Here, and throughout the paper, we use the formulas
\cite{abramowitz72}
\begin{equation}
\zeta(1-2n)=-\frac{B_{2n}}{2n},\quad \zeta(2n)=\frac{(2\pi)^{2n}}{2(2n)!}\,
|B_{2n}|,~~n=1,2,3,...,
\label{7.22}
\end{equation}
$B_{2n}$ being the Bernoulli numbers.

\subsection{Statistical-Mechanical Approach}
We recall the usual statistical mechanical expression for
the free energy,
\begin{subequations}
\bea
F&=&-kT\ln Z,\\
 \ln Z&=&-\sum_i\ln\left(1-e^{-\beta p_i}\right)\nonumber\\
&=&-\sum_{n=0}^\infty(n+1)^{d-2}\ln\left(1-e^{-\beta(n+1)/a}\right),
\label{freeen}
\eea
\end{subequations}
for conformally coupled scalars in $S^{d-1}$. Here the zero-point energy
has been subtracted.
 The latter, of course, is easily calculated,
\be
E_0=\sum_{n=0}^\infty(n+1)^{d-2}\frac{n+1}{2a}=\frac1{2a}\zeta(1-d).
\ee
The specific results for two and four dimensions are
\begin{subequations}
\bea
d=2:\quad&&E_0=-\frac1{2a}\frac12B_2=-\frac1{24a},\\
d=4:\quad&&E_0=-\frac1{2a}\frac14B_4=\frac1{240a}.\label{2.18}
\eea
\end{subequations}

For the temperature dependence, we differentiate the partition
function,
\be
E=
-\frac{d}{d\beta}\ln Z=\frac1a\sum_{n=1}^\infty\frac{n^3}{e^{2\pi n\delta}-1},
\quad\delta=\frac{\beta}{2\pi a}.
\label{smrep}
\ee
This is a very different representation from Eq.~(\ref{klft}).  Nevertheless,
from it we may obtain the same result we found above
if we use the Euler-Maclaurin sum formula,
\bea
\sum_{n=0}^\infty f(n)&=&\int_0^\infty dn\,f(n)+\frac{1}{2}[f(\infty)+f(0)]
\nonumber\\
&&\quad\mbox{}+\sum_{k=1}^\infty\frac1{(2k)!}B_{2k}\left[f^{(2k-1)}(\infty)
-f^{(2k-1)}(0)\right].
\eea
Doing so here yields for the integral
\be
\frac1a\int_0^\infty dn\frac{n^3}{e^{\beta n/a}-1}=\frac{a^3\pi^4}{15\beta^4},
\ee
while the third derivative term gives
\be
\frac3a\frac{B_4}{4!}=-\frac1{240a},
\ee
so we exactly reproduce the results of Kutasov and Larsen \cite{kl}, but
derived by a much more transparent method,
\be
E=U-E_0\sim\frac1{240 a}\left(\frac1{\delta^4}-1\right),\quad \delta\ll1,
\ee
coinciding with Eqs.~(\ref{sl}) and (\ref{2.18}).

\subsection{Comparison with Usual Finite-Temperature Casimir Effect}

The same argument could be used for a scalar field subject to Dirichlet
boundary conditions on parallel plates, with $d$ transverse
dimensions (see Ref.~\cite{mono}):
\bea
F&=&TV\int\frac{d^dk}{(2\pi)^{d+1}}\frac{\pi}{a}\sum_{n=-\infty}^\infty
\ln\left(1-e^{-\beta\sqrt{k^2+n^2\pi^2/a^2}}\right).
\eea
If we use the E-M formula, and replace the sum by an integral, and
rescale variables, $\beta k\to k$, $\beta \pi n/a\to n$, we obtain
\bea
F&=&\frac{TV}{\beta^{d+1}}\int\frac{d^dk\,dn}{(2\pi)^{d+1}}
\ln\left(1-e^{-\sqrt{k^2+n^2}}\right)\nonumber\\
&=&\frac{TV}{\beta^{d+1}}\frac{A_{d+1}}{(2\pi)^{d+1}}\int_0^\infty dk\,
k^d\ln\left(1-e^{-k}\right)\nonumber\\
&=&-VT^{d+2}\frac{\Gamma(d/2+1)\zeta(d+2)}{\pi^{d/2+1}},
\eea
where $A_d=2\pi^{d/2}/\Gamma(d/2)$ is the area of a $d$-dimensional sphere.
This result is not the correct linear high-temperature $T$ dependence,
but instead its negative appears as a (subdominant) term in
{\it low-temperature limit!}  (See, for example, 
Ref.~\cite{mono}, Eq.~(2.91).) What is different about this case is that there
is an inside (between the plates) and an outside, and the Stefan's law term,
being proportional to the volume, cancels between the two regions---it can
give no force on the plates.  Here there is no outside, so the volume term
must be included.  As explained in Ref.~\cite{mono}, p.~56, the high temperature
dependence for parallel plates of area $A$ is linear,
\begin{subequations}
\be
d=2:\quad \frac{F}{A}=-T\frac{\zeta(3)}{16\pi a^2},\quad aT\gg1,
\ee
while the low-temperature correction is cubic,
\be
d=2:\quad \frac{F}{A}
=-\frac{\pi^2}{1440 a^3}-\frac{\zeta(3)}{4\pi}T^3
+\frac{\pi^2 a}{90}T^4,\quad aT\ll1.
\ee
\end{subequations}
(Note that the cubic term does not contribute to the force on the plates,
being independent of $a$, but does contribute to the entropy,
$S\sim \frac{3\zeta(3)}{4\pi}T^2$, $aT\ll1$.)
\subsection{Relation Between Representations}
\label{sec:reps}
We have two representation for the Casimir energy, the one obtained
from the Green's function, Eq.~(\ref{klft}), and the one obtained from the
partition function, Eq.~(\ref{smrep}).  The relation between the two
can be found from the Poisson sum formula.  If the Fourier transform of
a function $b(x)$ is defined by
\be
c(\alpha)=\int_{-\infty}^\infty\frac{dx}{2\pi}e^{-i\alpha x}b(x),
\ee
then
\be
\sum_{n=-\infty}^\infty b(n)=2\pi\sum_{n=-\infty}^\infty c(2\pi n).
\label{psf}
\ee
This is just a rewriting of the elementary identity
\be
\sum_{n=-\infty}^\infty e^{-i2\pi nx}=\sum_{n=-\infty}^\infty\delta(x-n).
\ee
So start from Eq.~(\ref{smrep}) and consider
\be
b(x)=\frac1a\left\{\begin{array}{cc}
\frac{x^3}{e^{\beta x/a}-1},&x\ge0,\\
0,&x\le0.\end{array}\right.
\ee
It is easily seen that the Fourier transform is
\be
c(\alpha)=\frac1{2\pi a}
\sum_{k=0}^\infty\frac{\Gamma(4)}{[\beta(k+1)/a+i\alpha]^4}
=\frac1{2\pi a}
\left(\frac{a}{\beta}\right)^4\Gamma(4)\zeta\left(4,1+i\alpha a/\beta\right),
\ee
in terms of the Hurwitz zeta function.  Thus the energy is
\be
E=\frac1a\left(\frac{a}{\beta}\right)^4\Gamma(4)\sum_{n=-\infty}^\infty
\sum_{k=0}^\infty\frac1{\left(1+k-ia2\pi n/\beta\right)^4}.
\ee
If we sum this on $n$ first, we obtain the alternative expression
\be
E=\frac1a\Gamma(4)\frac1{(2\pi)^4}\left\{
\sum_{k=-\infty}^\infty\zeta(4,1-i\beta k
/(2\pi a))+\zeta(4)\left[\left(\frac{2\pi a}{\beta}\right)^4-1\right]\right\}.
\ee
By comparing with the original expression (\ref{smrep}) for $E$
we can replace the sum over Hurwitz zeta functions by the expression 
(\ref{klft}):
\bea
E=\frac1a\left(\frac{2\pi a}{\beta}\right)^4\sum_{n=1}^\infty
\frac{n^3}{e^{4\pi^2 an/\beta}-1}
+\frac1{240a}\left[(2\pi aT)^4-1\right],
\label{relation}
\eea
to which must be added the zero-point energy, which cancels the $T$-independent
term.  (Thus the $T=0$ term, which is the ZPE, arises from the sum,
again as directly verified by the E-M formula, and in the low temperature
limit the Stefan's law term is cancelled.)
Thus we again reproduce Eq.~(\ref{klft}).  The two representations are
adapted explicitly for the two limits:
\begin{subequations}
\bea
U&=&\frac1{240a}+\frac1a\sum_{n=1}^\infty\frac{n^3}{e^{2\pi n\delta}-1}\\
&=&\frac1{240a}(2\pi aT)^4+\frac1a(2\pi a T)^4\sum_{n=1}^\infty\frac{n^3}
{e^{2\pi n a/\delta}-1}.
\eea
\end{subequations}

\section{$d=2$ Conformal Scalar}
\label{sec:d2scalar}
Because there is a subtle issue involving zero-modes here, it is useful
to repeat the above calculation for $d=2$.  The partition function is given
by
\be
\ln Z=-\sum_{n=0}^\infty \ln\left(1-e^{-\beta(n+1)/a}\right).
\ee
This immediately gives the low-temperature representation,
\be
U=-\frac1{24a}+\frac1a \sum_{n=1}^\infty\frac{n}{e^{2\pi n\delta}-1},
\label{ltemplimd2scalar}
\ee
displaying an exponentially small correction to the zero-point energy if
$\beta\gg1$.  Again, by use of the Euler-Maclaurin sum formula we can obtain
the high-temperature limit,
\be
U\sim\frac1{24a}(2\pi a T)^2-\frac12 T,\quad aT\gg1.
\label{htemplimd2scalar}
\ee
This again coincides with the result found in Ref.~\cite{kl}.  However,
the linear term in $T$ is omitted in the analysis of Klemm et al.~\cite{klemm}
because they believe that the result can only be trusted in the limit
$T\to\infty$ \cite{siopsis}.  This omission is necessary for
their derivation of the Cardy formula \cite{cardy}.

It is easy to reproduce this result from the Green's function method.
After the finite-temperature substitutions, the expression is
\bea
U&=&\frac{i}{4\pi}\frac{8i\pi}{\beta}\sum_{n=0}^\infty{}'\left(\frac{2\pi n}
{\beta}\right)\frac{a\pi}{2}\left[1+\frac2{e^{4\pi^2an/\beta}-1}\right]
\nonumber\\
&=&\frac1{24a\delta^2}-\frac12 T-\frac1{a\delta^2}\sum_{n=1}^\infty\frac{n}
{e^{2\pi n/\delta}-1},
\label{general2drep}
\eea
which gives the explicit exponential corrections to the high temperature
limit (\ref{htemplimd2scalar}).
The low-temperature limit displayed in Eq.~(\ref{ltemplimd2scalar}) may
be easily obtained from this by using the Euler-Maclaurin sum formula.

A proof of the equivalence of the two representations (\ref{general2drep})
and (\ref{ltemplimd2scalar}) can be carried out along the lines sketched in
Sec.~\ref{sec:reps}.  
There is one complication, due to the presence of the zero-mode at $n=0$.  
The Poisson sum formula (\ref{psf}), applied to a function $b(x)$ which is positive
for $x\ge0$ and zero for $x<0$ implies a Fourier transform $c(\alpha)$ which
has poles in the upper half plane and hence implies $b(x)=0$ for $x\le0$.
This contradiction at $x=0$ is resolved by the following prescription:
Write
\be
\sum_{n=1}^\infty b(n)=\sum_{n=0}^\infty{}'b(n)-\frac12b(0), 
\ee
replace the latter sum by the sum over the Fourier transform $c(2\pi n)$,
and then when we interchange the sums and transform back the zero mode is
not present.  Thus only the term 
explicitly displayed above arises from the $n=0$ term.
In this way the equivalence between the two representations
(\ref{general2drep}) and (\ref{ltemplimd2scalar}), with the $-\frac12
T$ term, is established.

\section{Vector Field}
\label{sec:IV}
The analysis proceeds similarly to that given in Sec.~\ref{sec:II}.
For $S^{d-1}$ the degeneracy and eigenvalues are
\bea
D_l&=&\frac{2l\left(l+\frac{d}2-1\right)(l+d-2)(l+d-4)!}{(d-3)(l+1)!},\\
M_l^2&=&l(l+d-2),
\eea
so for $d=4$ if we again add the conformal coupling value 1 to $M_l^2$
we obtain for the Green's function mode sum
\be
\sum_{l=0}^\infty \frac{2l(l+2)}{(l+1)^2/a^2-\omega^2}\to
-\frac1{\omega^2}+i\pi a^2\left(\omega a-\frac1{\omega a}\right)
\left(1+\frac2{e^{-2\pi i\omega a}-1}\right).
\ee
After making the finite temperature replacement, we carry out the sum on
$n$, with the result
\bea
U&=&\frac{(2\pi aT)^4}{120 a}-\frac1{12}\frac{(2\pi a T)^2}{a}\nonumber\\
&&\quad\mbox{}+T+\frac{2(2\pi a T)^2}{a}\sum_{n=1}^\infty\left[n+(2\pi a T)^2
n^3\right]\frac1{e^{4\pi^2aTn}-1},
\label{htexv}
\eea
where the $T$ term comes from the $n=0$ term in the sum. 
[Note that the leading term is exactly twice that found for a  scalar
field found in Eq.~(\ref{sl}), as we would expect.] Since the remaining
sum is exponentially small in the large $T$ limit, this form is well-adapted
for high temperature. 
(The $T^4$ and $T^2$ terms are as given in Ref.~\cite{kl},
while the linear term is implicitly given there.)
 However, it is exact, and by using the Euler-Maclaurin
sum formula it yields the low temperature limit,
\be
U\sim\frac{11}{120a},\quad aT\ll1,
\ee
up to exponentially small corrections.  The latter may be directly inferred
from the partition function,
\be
\ln Z=-\sum_{l=1}^\infty 2l(l+2)\ln\left(1-e^{-\beta(l+1)/a}\right).
\ee  By taking the negative derivative of this with respect to $\beta$ we
obtain the alternative representation
\be
U=\frac{11}{120a}+\frac2a\sum_{l=1}^\infty\frac{l(l^2-1)}{e^{\beta l/a}-1}.
\label{ltexv}
\ee
The Euler-Maclaurin formula applied to this last sum yields the leading
high-temperature term in Eq.~(\ref{htexv}), including the term linear in $T$,
and the exact equivalence of the two expressions (\ref{ltexv}) and
 (\ref{htexv}) again is demonstrated by the Poisson sum formula, including
the zero-mode prescription enunciated in Sec.~\ref{sec:d2scalar}.

\section{Weyl Fermions}
\label{sec:V}
The procedure is now routine, except for the complications due to fermions.  
The degeneracies and eigenvalues are
\be
D_l=2(l+2)(l+1),\quad M_l^2=l(l+3),
\ee
so recalling the minus sign associated with a fermionic trace,
and that the fermionic thermal Green's functions must be antiperiodic, we have
the following expression for the energy,
\bea
U&=&\frac{i}{4\pi}\frac{8\pi i}\beta 2a^2\sum_{n=0}^\infty\frac{i\pi a}2
\left[\left(\frac{i2\pi(n+1/2)}{\beta}\right)^3-\frac13\left(\frac{i2\pi
(n+1/2)}{\beta}\right)\right]\nonumber\\
&&\qquad\times\left(1+\frac4{e^{8\pi^2a(n+1/2)/\beta}-1}
-\frac2{e^{4\pi^2a(n+1/2)/\beta}
-1}\right)\nonumber\\
&=&\frac1a\bigg\{\frac7{960}\delta^{-4}-\frac1{96}\delta^{-2}\nonumber\\
&&\quad-\frac14\sum_{n=0}^\infty\left[(2n+1)^3\delta^{-4}+(2n+1)\delta^{-1}
\right]\left(\frac2{e^{2\pi(2n+1)/\delta}-1}-\frac1{e^{\pi(2n+1)/\delta}-1}
\right).
\label{htexf}
\eea
The low-temperature limit (the zero-point energy) may be obtained from
this by the Euler-Maclaurin formula, and the exponential
corrections in that limit may be obtained directly from the 
partition function,
\be
\ln Z=\sum_{n=1}^\infty 2n(n+1)\ln\left(1+e^{-\beta(2n+1)/2a}\right).
\ee
That is
\be
U=\frac1a\left[\frac{17}{960}+\sum_{n=1}^\infty \frac{n(n+1)(2n+1)}
{e^{\beta(2n+1)/2a}+1}\right].
\label{ltexf}
\ee
The equivalence between Eqs.~(\ref{htexf}) and (\ref{ltexf}) may be
demonstrated again either though the Euler-Maclaurin sum formula applied
to the latter, or exactly through the use of the Poisson sum formula.

\section{Entropy Bounds}
\label{sec:VI}
From the above results, thermodynamic information may extracted in terms of
the free energy (with zero-point energy $E_0$ subtracted),
\be
F=-T\ln Z, 
\ee
in terms of which the energy
\be
E\equiv U-E_0=-\frac\partial{\partial \beta}\ln Z=\frac\partial{\partial\delta}
\delta F,
\ee
and the entropy
\be
S=2\pi a\delta^2\frac{\partial}{\partial\delta}F=\beta(E-F)
\ee
may be extracted.

\subsection{Two-dimensional scalar}
\label{tdimscalar}
Klemm et al.~\cite{klemm} ignore the linear $T$ term in the energy,
and so have for $\delta\ll1$ (see Sec.~\ref{sec:d2scalar})
\begin{subequations}
\bea
E&=&\frac1{24a}(\delta^{-2}+1),\\
F&=&-\frac1{24a}(\delta^{-2}-1),\\
S&=&\frac\pi6\delta^{-1}.
\eea
\end{subequations}
These imply the Verlinde-Cardy formula \cite{verlinde,cardy}
\be
S=4\pi a\sqrt{-E_0(E+E_0)},
\ee
and the entropy bound
\be
\frac{S}{2\pi a E}=2\frac{\delta}{\delta^2+1}\le1.
\ee
However, this result is not meaningful as it stands.  Even in the 
high-temperature limit we must add the term linear in temperature to the 
energy, which implies instead from Eq.~(\ref{htemplimd2scalar}), for
$\delta\ll1$, that
\begin{subequations}
\bea
E&=&\frac1{24a}(\delta^{-2}+1)-\frac1{4\pi a\delta},\\
F&=&-\frac1{24a}(\delta^{-2}-1)-\frac1{4\pi a\delta}\ln\delta,\\
S&=&\frac\pi6\delta^{-1}+\frac12(\ln\delta-1).
\eea
\end{subequations}
The ratio of $S$ to $E$ is then unbounded as $\delta\to\infty$.  Yet this
takes us to the low-temperature regime, where we must use the leading
exponential corrections, for $\delta\gg1$,
\begin{subequations}
\bea
E&\sim&\frac1a e^{-\beta/a},\\
F&\sim&-\frac1\beta e^{-\beta/a},\\
S&\sim&\frac\beta{a} e^{-\beta/a},
\label{lowtemp2dentropy}
\eea
\end{subequations}
so the entropy-energy ratio is
\be
\frac{S}{2\pi a E}=\delta,\quad \delta\gg1.
\ee
It is apparent that this latter result is universal because the energy
always dominates the free energy in the low temperature regime.

\subsection{Entropy Bounds in Four Dimensions}
In the following we will consider cases with $N_s$ conformal scalars,
$N_v$ vectors, and $N_f$ Weyl fermions. For example, the ${\cal N}=4$ 
supersymmetric multiplet has $N_s=6$, $N_v=1$, and $N_f=4$. 

In the high-temperature regime we may write the free energy, energy, and
entropy as\footnote{What is called the Cardy formula is simply
the observation that the leading behavior of $S$ is the geometric
mean of the leading and subleading terms in $E$.  The term ``Casimir energy''
for the latter is misleading in other than $1+1$ dimensions.
The entire energy $U$ is due to quantum and thermal fluctuations, so it all
should properly be reckoned as Casimir energy.}
\begin{subequations}
\label{form}
\bea
F&\sim&-\frac1a[a_4\delta^{-4}+a_2\delta^{-2}+a_1\delta^{-1}\ln\delta+a_0],\\
E&\sim&\frac1a[3a_4\delta^{-4}+a_2\delta^{-2}-a_0-a_1\delta^{-1}],\\
S&\sim&2\pi[4a_4\delta^{-3}+2a_2\delta^{-1}-a_1(1-\ln\delta)].
\eea
\end{subequations}
Here the coefficients were determined in the previous sections to be
[see Eqs.~(\ref{relation}), (\ref{htexv}), and (\ref{htexf})]
\begin{subequations}
\bea
a_4&=&\frac{N_s}{720}+\frac{N_v}{360}+\frac{7N_f}{2880},\\
a_2&=&-\frac{N_v}{12}-\frac{N_f}{96},\\
a_0&=&3a_4-a_2,\\
a_1&=&-\frac{N_v}{2\pi}.
\eea
\end{subequations}
Even ignoring the $a_1$ term, Klemm et al.~\cite{klemm} note that no
entropy bound is possible, unless special choices are made for the field
multiplicities.  For the ${\cal N}=4$ case the first three coefficients are
\be
a_4=\frac1{48},\quad a_2=-\frac18,\quad a_0=\frac3{16},
\ee
and the entropy-energy ratio becomes
\be
\frac{S}{2\pi a E}=\frac{1-\ln\delta+\frac{\pi}6\delta^{-3}(1-3\delta^2)}
{\delta^{-1}+\frac{\pi}8\delta^{-4}(1+\delta^2)(1-3\delta^2)}.
\label{enenratio}
\ee
If the $a_1$ terms here were omitted, the zero in both the energy and
entropy at $\delta^2=1/3$ would cancel, and we would have the limit
given in Ref.~\cite{klemm}:
\be
\frac{S}{2\pi a E}=\frac43\frac\delta{1+\delta^2}\le\frac43
\label{klemmratio}
\ee
in the high temperature regime.  But $a_1\ne0$, and the ratio (\ref{enenratio})
diverges as $\delta\to\infty$.  Of course that limit is the low-temperature one,
but the argument given in Sec.~\ref{tdimscalar} then applies and shows that
\be
\frac{S}{2\pi a E}\sim\delta,\quad \delta\to\infty.
\ee
Although in this limit both the entropy and the subtracted energy are
exponentially small, their ratio is unbounded.

The reader should note that we are not in formal disagreement with previous
studies \cite{kl,klemm}.  The interest there was restricted to high
temperature, which is presumably all that is relevant to nearly the
entire history
of the universe.\footnote{It might be noticed that before or
after photon decoupling, but after inflation,
the value of $aT$ stays nearly constant \cite{Weinberg},
essentially reflecting entropy conservation.  That value
is far into the high temperature regime, the present value of $\delta$
being $\delta_0\sim 10^{-30}$.  Insofar as it is permissible to speak
of temperature during inflation, $aT$ is also constant then, but of a
much smaller value, which value increases dramatically during reheating.} 
  In that case, only the leading terms in $1/\delta$ are
relevant, and the ratio of entropy to energy is always of order $\delta\ll1$.
It is not surprising that such results as Eq.~(\ref{klemmratio}) are
an unreliable guide to the moderate and low temperature regimes,
which might be relevant in the very earliest 
(pre-inflationary) stages of the universe.

Another point, which is more closely connected with physics, is that it is 
permissible to make use of the thermodynamical formalism for fluctuating 
quasi-classical systems only when the temperature $T$ is sufficiently high.  
As discussed recently by Das et al.~\cite{das}, their Eq.~(12), one must 
have
\begin{equation}
T\gg \frac{1}{\tau},
\end{equation}
where $\tau$ is the relaxation time, in order to use thermodynamics. Now, 
in our case it is presumably legitimate to estimate $\tau$ to be of the 
same order of magnitude as the transit time for light across a distance 
of order $a$, i.e., $\tau \sim a$. This leads to the condition $T \gg 1/a$, 
which actually means
\begin{equation}
\delta \ll 1.
\end{equation}
In other words, the physical condition for using Eq.~(\ref{form}) 
above seems to be  
that the linear term, 
which is of  order $\delta^3$ relative to the first term, 
is negligible!  It seems that we must be careful in not assigning too much 
physical significance to the subleading corrections.

\section{On the half Einstein universe}
\label{sec:VII}

The metric of the static Einstein universe has continuously attracted 
interest, both because it is easily tractable analytically and also because 
it is conformally equivalent to all the closed Robertson-Walker metrics. 
The Einstein metric can be written as
\begin{equation}
ds^2= -dt^2+a^2[d\chi^2+\sin^2 \chi \,(d\theta^2+\sin^2\theta \,d\phi^2)],
\label{7.1}
\end{equation}
where $\theta \in [0, \pi]$ and $\phi \in [0,2\pi]$. In the 
case of the full Einstein universe, $\chi \in [0,\pi]$. The energy 
density $\rho$ consists of two parts, one a matter (dust) part, $\rho_0$, and 
one a vacuum part, $\Lambda/8\pi G$, $\Lambda$ being the cosmological constant. 
The pressure is $p=-\Lambda/8\pi G$. From the Friedmann equations, 
$\rho_0=\Lambda/4\pi G$, and the scale factor becomes 
$a=\Lambda^{-1/2}=(4\pi G\rho_0)^{-1/2}$. 

The {\it half} Einstein universe, in which $\chi$ varies only from 0
 to $\pi/2$, turns out to be an interesting variant of the Einstein 
static universe idea; see, for instance, Refs.~\cite{kennedy80} and 
\cite{bayin93}. As discussed in Ref.~\cite{brevik97}, we can consider 
this universe as a three-dimensional spherical volume, spanned by the 
``radius" $\chi$ and the angular coordinates $\theta$ and $\phi$, 
closed by a two-dimensional spherical surface lying at $\chi=\pi/2$. Let 
us assume that this surface is perfectly conducting. This is the simplest 
imaginable option in the electromagnetic case, and is analogous to the 
Dirichlet boundary condition in scalar field theory. The assumption 
allows the consideration of standing electromagnetic waves in this spherical 
cavity. We expect to find the same possibility of dividing the possible 
eigenmodes into independent TE modes and TM modes, as we do when dealing 
with the conventional standing modes in the Minkowski metric. Explicit 
calculation of the fundamental electromagnetic field modes was actually given 
in Ref.~\cite{brevik97}.

Let us assume an orthonormal basis, designated by carets, 
$\{ \omega^{\hat{t}}, \omega^{\hat{\chi}}, 
\omega^{\hat{\theta}},\,\omega^{\hat{\phi}}\}=\{dt, a\,d\chi, \,
a\sin \chi d\theta,\, a\sin\chi \sin\theta \,d\phi \}$, and split off 
the time factor as $e^{-i\omega t}$. From Maxwell's equations in the 
curvilinear space we find the governing equation for $E_{\hat{\chi}}$ 
(or $H_{\hat{\chi}}$).
 Denoting these field components collectively by $X$, and writing 
$E_{\hat{\chi}}(\chi, \theta, \phi)=E_{\hat{\chi}}(\chi)Y_{l}^m
(\theta, \phi)$, we obtain
\begin{equation}
\frac{d^2}{d\chi^2}\left( \sin^2\chi\, X\right)
+(\omega a)^2\sin^2\chi\,X-l(l+1)X=0.
\label{7.2}
\end{equation}
The solution of this equation is known:
\begin{equation}
X \propto \sin^{l-1}\chi\, C_{n-l}^{(l+1)}(\cos\chi),
\label{7.3}
\end{equation}
where $n$ is an integer,  $~C_{n-1}^{(l+1)}$ being the Gegenbauer 
polynomials \cite{abramowitz72}. The differential equation satisfied by 
$C_p^{(\alpha)}(x), p \ge 1$ is
\begin{equation}
(1-x^2)\,{C_p^{(\alpha)}}''(x)-(2\alpha+1)x\,{C_p^{(\alpha)}}'(x)
+p(p+2\alpha)\,C_p^{(\alpha)}(x)=0.
\label{7.4}
\end{equation}
Inserting Eq.~(\ref{7.3}) into Eq.~(\ref{7.2}) we obtain the eigenfrequencies
\begin{equation}
\omega_n=\frac{n+1}{a}, \quad n\ge l \ge 1.
\label{7.5}
\end{equation}
For the electromagnetic field, we know that $l\ge 1$. Moreover, in 
order to avoid infinities at the origin $\chi=0$, we must have 
$n-l \ge 0$ in Eq.~(\ref{7.3}) \cite{brevik97}.

As mentioned above, the TE and TM modes are independent and can be 
considered separately. In the following, we consider the TE modes  only.

\subsection{The TE Modes}

The electromagnetic boundary condition at the conducting surface is in this case
\begin{equation}
H_{\hat{\chi}}=0, \quad \chi=\pi/2.
\label{7.6}
\end{equation}
Since $l\ge 1$ the condition (\ref{7.6}), 
together with the general property that 
$C_p^{(\alpha)}(0)=0$ when $p$ is odd \cite{brevik97}, shows that the 
subscript $(n-l)$ in Eq.~(\ref{7.3}) must be odd. The eigenvalues 
(\ref{7.5}) are seen to 
depend only on $n$, so that they are degenerate with respect to $l$ as 
well as to the magnetic quantum number $m$ appearing in $Y_{l}^m$. We 
accordingly have to sum $(2l+1)$ over all the admissible values of $l$.

Assume first that $n$ is even. According to the above, $l$ can then take 
all odd values between 1 and $(n-1)$, so that the degeneracy becomes
\begin{subequations}
\begin{equation}
\sum_{l=1,3,5,..(n-1)}(2l+1)=\frac{n}{2}(n+1),\quad n=2,4,6,\dots.
\label{7.7}
\end{equation}
Next, if $n$ is odd, $l$ can take all even values between 2 and $(n-1)$. 
This leads to the degeneracy
\begin{equation}
\sum_{l=2,4,6,..(n-1)}(2l+1)=\frac{n-1}{2}(n+2),\quad n=1,3,5,\dots.
\label{7.8}
\end{equation}
\end{subequations}
The partition function for the TE modes is accordingly
\begin{eqnarray}
 \ln Z^{\rm TE}&=&-\sum_{n=2,4,6,..}^\infty \frac{n}{2}(n+1)\ln 
\left( 1-e^{-\beta (n+1)/a} \right)\nonumber\\
&&\quad\mbox{}-\sum_{n=1,3,5,..}^\infty \frac{n-1}{2}(n+2)\ln 
\left( 1-e^{-\beta (n+1)/a}\right)\nonumber\\
&=&-\sum_{n=1}^\infty n(2n+1)\ln \left( 1-e^{-\beta (2n+1)/a}\right)\nonumber\\
&&\quad\mbox{}-\sum_{n=1}^\infty (n-1)(2n+1)\ln 
\left( 1-e^{-2\beta n/a}\right).
\label{7.9}
\end{eqnarray}
We write this expression as a sum of two terms,
\begin{equation}
\ln Z^{\rm TE}=A+B,
\label{7.10}
\end{equation}
where $A$, the first term in Eq.~(\ref{7.9}), is rewritten as
\begin{equation}
A=-\frac{1}{2}\sum_{n=1}^\infty (2n+1)^2\ln 
\left(1-e^{-2\pi (2n+1)\delta} \right)
+\frac{1}{2}\sum_{n=1}^\infty (2n+1)\ln \left(1-e^{-2\pi (2n+1)\delta}\right),
\label{7.11}
\end{equation}
where we recall $\delta=\beta/(2\pi a)$. The generic form is thus
\begin{equation}
G^{(\alpha)}=\sum_{n=1}^\infty (2n+1)^\alpha \ln 
\left( 1-e^{-2\pi (2n+1)\delta}\right),\quad \alpha=2,1.
\label{7.12}
\end{equation}

For variety's sake, we evaluate this expression using a variation on the 
method given in Ref.~\cite{kl}. (We also used this method to derive
the high temperature results in Sec.~\ref{sec:II}.)
Expanding the logarithm and thereafter differentiating with  
respect to $\delta$, we have
\begin{equation}
\frac{\partial G^{(\alpha)}}{\partial \delta}=2\pi 
\sum_{n=1}^\infty (2n+1)^{\alpha+1}\sum_{k=1}^\infty e^{-2\pi k(2n+1)\delta},
\label{7.13}
\end{equation}
which can be processed further by inserting
\begin{equation}
e^{-x}=\frac{1}{2\pi i}\int_C ds\, x^{-s}\,\Gamma(s).
\label{7.14}
\end{equation}
The integration is here taken along a line parallel to the imaginary axis 
with sufficiently large value of $\Re\, s$. By introducing Riemann's 
zeta function $\zeta(s)$ via the sum over $k$ we obtain
\begin{equation}
\frac{\partial G^{(\alpha)}}{\partial \delta}=\frac{1}{i}\sum_{n=1}^\infty 
(2n+1)^{\alpha+1}\int_Cds\, [2\pi (2n+1)\delta]^{-s}\,\Gamma(s)\zeta(s).
\label{7.15}
\end{equation}
The sum over $n$ leads to 
\begin{equation}
\sum_{n=1}^\infty (2n+1)^{-s+\alpha+1}=(1-2^{-s+\alpha+1})\zeta(s-\alpha-1)-1,
\label{7.18}
\end{equation}
and the expression (\ref{7.15}) can conveniently be written as
\begin{equation}
\frac{\partial G^{(\alpha)}}{\partial \delta}=\{I \}^{(\alpha)}+\{II\},
\label{7.19}
\end{equation}
with
\begin{subequations}
\begin{eqnarray}
\{I\}^{(\alpha)}&=&\frac{1}{i}\int_Cds\,(2\pi \delta)^{-s}
(1-2^{-s+\alpha+1})\zeta(s-\alpha-1)\Gamma(s)\zeta(s),\label{7.20}
\\
\{II\}&=&-\frac{1}{i}\int_C ds\,(2\pi \delta)^{-s}\Gamma(s)\zeta(s).
\label{7.21}
\end{eqnarray}
\end{subequations} 
Consider first the expression (\ref{7.20}). Because of the poles of the zeta 
functions it is properly defined only for $\Re\,s >\alpha +2$ 
($\alpha>0$ assumed). For lower values of $s$, we define the zeta 
functions in terms of their analytic continuations. The integrand has 
poles at $s=1$ and $s=\alpha +2$ from the zeta functions, and at 
$s=-n,~n=0,1,2,...$ from the gamma function. There are, however, zeroes 
from the zeta functions, at $s=-2n$ and $s=\alpha+1-2n$, $n=1,2,3,...,$ 
so that the remaining number of poles becomes effectively reduced.

For $\alpha=2$, there remain poles only at $s=4$ and $s=0$. 
We distort the contour to encircle these poles and find, by means of the 
formulas (\ref{7.22}), that
\begin{equation}
\{I\}^{(2)}=\frac{\pi}{240\delta^4}+\frac{7\pi}{120}.
\label{7.23}
\end{equation}
For $\alpha=1$, the expression  (\ref{7.20}) 
has poles at $s=3,1,-1,-3,...$. Only 
$s=3,1$ need to be taken into account since we assume that $\delta \ll 1$ 
(high temperatures), implying that terms of order $\delta$ are negligible. 
We get
\begin{equation}
\{I\}^{(1)}=\frac{\zeta(3)}{4\pi^2\delta^3}+\frac{1}{12\delta}.
\label{7.24}
\end{equation}
Consider now the expression $\{II\}$, Eq.~(\ref{7.21}), 
which is independent of 
$\alpha$. The integrand has poles at  $s=1,-1,-3,-5,...$.  Because of the 
smallness of $\delta$ we have to include only the lowest value, $s=1$. We get
\begin{equation}
\{II\}=-\frac{1}{\delta}.
\label{7.25}
\end{equation}
It actually turns out that when constructing $\partial A/\partial \delta$ 
according to Eqs.~(\ref{7.11}), (\ref{7.12}), and (\ref{7.19}) 
the quantity $\{II\}$ does not contribute:
\bea
\frac{\partial A}{\partial \delta}&=&-\frac{1}{2}\frac{\partial G^{(2)}}
{\partial\delta}
+\frac{1}{2}\frac{\partial G^{(1)}}{\partial\delta}
=-\frac{1}{2}\{I\}^{(2)}+\frac{1}{2}\{I\}^{(1)}\nonumber\\
&=&-\frac{\pi}{480\delta^4}+\frac{\zeta(3)}{8\pi^2\delta^3}+\frac{1}{24\delta}
-\frac{7\pi}{240}.
\label{7.26}
\eea

We now consider the second term $B$ in Eq.~(\ref{7.10}):
\begin{equation}
B=-\sum_{n=1}^\infty (2n^2-n-1)\ln\left( 1-e^{-4\pi n\delta}\right),
\label{7.27}
\end{equation}
whose generic form is
\begin{equation}
H^{(\alpha)}=\sum_{n=1}^\infty n^\alpha \ln \left( 1-e^{-4\pi n\delta}\right),
\quad \alpha=2,1,0.
\label{7.28}
\end{equation}
A calculation along the same lines as above yields
\begin{equation}
\frac{\partial H^{(\alpha)}}{\partial \delta}=\frac{2}{i}\int_Cds\,
(4\pi\delta)^{-s}\,\zeta(s-\alpha-1)\Gamma(s)\zeta(s),
\label{7.29}
\end{equation}
and we get (the values of $s$ giving poles are written within the parentheses)
\begin{subequations}
\begin{eqnarray}
\frac{\partial H^{(2)}}{\partial \delta}&=&\frac{\pi}{960\,\delta^4}
-\frac{\pi}{60} \quad(s=4,0),\label{7.30}\\
\frac{\partial H^{(1)}}{\partial \delta}&=&
\frac{\zeta(3)}{8\pi^2\delta^3}-\frac{1}{12\,\delta}\quad (s=3,1),\label{7.31}
\\
\frac{\partial H^{(0)}}{\partial \delta}&=&
\frac{\pi}{24\,\delta^2}-\frac{1}{2\delta}+\frac{\pi}{6}\quad(s=2,1,0).
\label{7.32}
\end{eqnarray}
\end{subequations}
Thus
\begin{equation}
\frac{\partial B}{\partial \delta}=
-\frac{\pi}{480\,\delta^4}+\frac{\zeta(3)}{8\pi^2\delta^3}
+\frac{\pi}{24\,\delta^2}-\frac{7}{12\,\delta}+\frac{\pi}{5}.
\label{7.33}
\end{equation}
Adding Eqs.~(\ref{7.26}) and (\ref{7.33}) 
we thus get for the thermodynamic energy $E^{\rm TE}$
of the TE modes
\begin{eqnarray}
 2\pi a E^{\rm TE}=-\frac{\partial}{\partial \delta}\ln Z^{\rm TE}
= \frac{\pi}{240\,\delta^4}-\frac{\zeta(3)}{4\pi^2\delta^3}-
\frac{\pi}{24\,\delta^2}+\frac{13}{24\,\delta}-\frac{41\pi}{240}.
\label{7.34}
\end{eqnarray}
This corresponds to the free energy
\begin{eqnarray}
 \beta F^{\rm TE}=-\ln Z^{\rm TE} 
=-\frac{\pi}{720\,\delta^3}+\frac{\zeta(3)}{8\pi^2\delta^2}
+\frac{\pi}{24\,\delta}+\frac{13}{24}\ln \delta -\frac{41\pi}{240}\,\delta.
\label{halfeinresult}
\end{eqnarray}
The same result is obtained in Appendix \ref{app:B} 
using the Euler-Maclaurin method
explained in Sec.~\ref{sec:II}.

\subsection{Comment on TM Modes}
Although we have not considered the TM modes, we expect on physical grounds 
that they should contribute  the same amount to the leading term in the 
energy as do the TE modes. Accordingly, for the total thermodynamic energy 
$U=U^{\rm TE}+U^{\rm TM}$ in the half Einstein universe we expect
\begin{equation}
2\pi aU=\frac{\pi}{120\,\delta^4}+{\rm subleading~~terms}.
\end{equation}
 This is the same leading term expression as obtained for a conformal scalar
field in the $R\times S^3$ geometry considered in Sec.~II, and half that for
a vector, see Eq.~(\ref{htexv}).
We conclude that in the high-temperature limit, the ratio of entropy to
energy is the same as given in Eq.~(\ref{klemmratio}),
\be
\frac{S}{2\pi a E}\sim\frac43\delta,\quad\delta\ll1.
\ee

\section{Conclusion}

There are two main points in the Verlinde proposal \cite{verlinde}.
First, the Friedmann-Robertson-Walker (FRW) 
equation may be written formally very
similarly to the Cardy entropy equation \cite{cardy}. Even 
in the presence of higher derivative
gravitational terms, or unusual matter content, or in the presence of
quantum corrections, one can redefine the energy density to arrive at the
standard form of the  FRW equations.

Second, the free energy for higher dimensional conformal field theory
has formally 
(at least for high temperature) the same structure
 as in the two-dimensional Cardy 
case.  Moreover, assuming that (a closed) universe has extensive and
subextensive (``Casimir'') contributions to its energy and entropy it was 
suspected \cite{verlinde} that the ``Casimir energy''  is less than 
the Bekenstein-Hawking energy. As a result, the 
entropy/energy cosmological bound appears.
It is, however,  clear that this bound cannot be universal.
For example, quantum corrections \cite{odintsov} or the choice of
an unusual state of matter would significantly alter such a bound.

Our main qualitative result in this paper is that entropy/energy bounds
should be relevant only in the ultra-high temperature limit, which
applies to the universe after inflation.
With the decrease of temperature the bound becomes much less 
reliable, until at low temperature ($aT\ll1$),
which might be the case in the very early universe,
the entropy dominates the energy.
This effect occurs already for conformal matter. As is apparent
from the results of 
Appendix \ref{app:A} the situation for non-conformal matter is much more 
complicated. Hence, it is unclear if entropy/energy bounds should exist 
at all even for high temperature. This question will be discussed 
elsewhere.

\begin{acknowledgments}
The work of KAM was supported in part by a grant from the US Department of
Energy.  He further thanks Yun Wang for helpful discussion.
\end{acknowledgments}

\appendix
\section{Conformal Symmetry Breaking}
\label{app:A}
In this Appendix, we
wish to consider conformal symmetry breaking by choosing for the
eigenvalue of the scalar field
\be
M_l^2=l(l+2)+\chi=(l+1)^2+\chi-1.
\ee
The conformal value considered Sec.~\ref{sec:II}
 corresponded to $\chi=1$.  As we shall
see, the analysis is considerably more subtle when $\chi\ne1$.

Let us first attempt to
calculate the zero-point energy, the Casimir energy at zero
temperature.  Using Eq.~(\ref{ce}) and carrying out the sum according to
Eq.~(\ref{cotan}), we have, after dropping the constant,
\be
E_0=\frac{i}8\int_c d\omega\,\omega^2a^2\alpha\cot\pi\alpha,
\ee
where
\be
\alpha=\sqrt{\omega^2a^2+1-\chi}.
\ee
Remember that $c$ encloses the poles on the positive real axis in a clockwise
sense, and those on the negative real axis in a counterclockwise sense.
There is no pole at $\alpha=0$, but poles at $\alpha=\pm1,\pm2,\dots$.
Changing to the variable $\alpha$, we have
\be
E_0=\frac1{4a}\int_{c'}\alpha^2d\alpha\sqrt{\alpha^2+\chi-1}
\left(\frac1{e^{-2\pi i\alpha}-1}+\frac12\right).
\label{A4}
\ee
Here we draw branch lines on the positive and negative real axes, starting
from the branch points at $\alpha=\pm\sqrt{1-\chi}$, respectively, and $c'$
encircles those branch lines in a $\mp$ sense respectively.

Now we see a serious problem.  When $\chi=1$ there is no branch line, and
the 1/2 term goes away, as it is not singular.  But with $\chi\ne1$ 
each individual term possesses a branch line (although the entire
integrand in Eq.~(\ref{A4}) does not), and this expression seems irreducibly
divergent.

Nevertheless, let us press on a bit.  The finite temperature version of
this is
\bea
U&=&\frac\pi\beta\sum_{n=1}^\infty{}'\left(\frac{2\pi an}{\beta}\right)^2\alpha
\cot\pi\alpha\nonumber\\
&=&\frac\pi\beta\sum_{n=1}^\infty\left(\frac{2\pi an}{\beta}\right)^2
\sqrt{\left(\frac{2\pi an}{\beta}\right)^2+\chi-1}\left[1+\frac2{e^{2\pi\sqrt{(2\pi
n a/\beta)^2+\chi-1}}-1}\right].
\eea
Now we have further difficulties.
What is one to make of the leading divergent term? Also, the
square root is not real if $a/\beta
\ll1$. Ambiguities also abound if we start from the partition function,
and try to derive from it the high-temperature limit.
It seems apparent that conformal symmetry breaking poses problems for
Casimir calculations that have not yet been solved.  This is not a new
observation, of course.  For example, it is known \cite{mass} that a
massive field in flat space bounded by a spherical shell possesses divergences
that are not present for massless fields.

\section{Euler-Maclaurin Method}
\label{app:B}
In this Appendix we sketch the derivation of the result (\ref{7.34})
using the Euler-Maclaurin method given in Sec.~II.  We write the result for
the subtracted energy 
\begin{equation}
E=-\frac{\partial}{\partial\beta}\ln Z^{\rm TE}=\frac1a\sum_{n=1}^\infty
[f(n)+g(n)],
\end{equation}
where
\begin{eqnarray}
f(n)&=&\frac{n(2n+1)^2}{e^{\beta(2n+1)/a}-1},\\
g(n)&=&\frac{n(2n+3)2(n+1)}{e^{2(n+1)\beta/a}-1}.
\end{eqnarray}
To get the high-temperature limit, we use the Euler-Maclaurin sum formula.
The first integral term is
\begin{equation}
\int_0^\infty dn\,f(n)=\frac14\left(\frac{a}{\beta}\right)^4
\int_{\beta/a}^\infty du\frac{u^2(u-\beta/a)}{e^u-1}.
\end{equation}
For $\beta/a\ll1$ this is easily seen to be
\begin{equation}
\int_0^\infty dn\,f(n)=\frac{\pi^4}{60}\left(\frac{a}{\beta}\right)^4
-\frac{\zeta(3)}2\left(\frac{a}{\beta}\right)^3+\frac1{24}\frac{a}{\beta}-
\frac1{96}.
\end{equation}
In the same way, the integral of the second function is seen to be
\begin{equation}
\int_0^\infty dn\,g(n)=\frac{\pi^4}{60}\left(\frac{a}{\beta}\right)^4
-\frac{\zeta(3)}2\left(\frac{a}{\beta}\right)^3-\frac{\pi^2}{12}\left(
\frac{a}{\beta}\right)^2
+\frac{5}{6}\frac{a}{\beta}-\frac1{3}.
\end{equation}
The contribution of these two integrals to the energy is
\begin{equation}
E_a=\frac1a\left[\frac{\pi^4}{30}\left(\frac{a}{\beta}\right)^4
-\zeta(3)\left(\frac{a}{\beta}\right)^3-\frac{\pi^2}{12}
\left(\frac{a}{\beta}\right)^2+\frac78\left(\frac{a}{\beta}\right)
-\frac{11}{32}\right].
\label{ea}
\end{equation}
The derivative terms in the Euler-Maclaurin formula may be obtained from the
small $\beta$ forms:
\begin{eqnarray}
f(n)&=&\frac{a}{\beta}n(2n+1)-\frac12 n(2n+1)^2+{\cal O}\left(\frac\beta{a}
\right),\\
g(n)&=&\frac{a}{\beta}n(2n+3)-n(2n+3)(n+1)+{\cal O}\left(\frac\beta{a}
\right).
\end{eqnarray}
Thus, for high temperature, the only nonzero derivatives appearing are
\begin{eqnarray}
f(0)&=&0,\quad f'(0)=\frac{a}{\beta}-\frac12,\quad  f'''(0)=-12,\\
g(0)&=&0,\quad g'(0)=3\frac{a}{\beta}-3,\quad  g'''(0)=-12.
\end{eqnarray}
These give a contribution to the energy of
\begin{eqnarray}
E_b&=&\frac1a\left[-\frac{B_2}{2}\left(\frac{4a}{\beta}-\frac72\right)
-\frac{B_4}{24}\left(-24\right)\right]\nonumber\\
&=&\frac1a\left(-\frac{a}{3\beta}+\frac{31}{120}\right).
\end{eqnarray}
Combining this with $E_a$, Eq.~(\ref{ea}), we obtain exactly the result
found in Sec.~VII, Eq.~(\ref{7.34}):
\begin{equation}
E=\frac1a\left[\frac1{480\delta^4}-\frac{\zeta(3)}{8\pi^3\delta^3}
-\frac1{48\delta^2}+\frac{13}{48\pi\delta}-\frac{41}{480}\right].
\end{equation}

\section{Zero-Mode Contribution}

There are two approaches to the investigation of vacuum (free) energy 
when the eigenvalue problem is well-posed.
In the first, one drops the zero-mode contribution. In usual quantum
field theory this can be justified
by the renormalization of the corresponding coupling constant (normally,
the cosmological constant). In the second, 
one keeps the zero-mode contribution 
explicitly. Clearly, both these approaches are merely conventions and 
one must supply some physical considerations to argue in favor of one
or the other.
 Moreover, the interpretation in terms of renormalization of 
coupling constants in quantum field theory at nonzero temperature becomes 
quite unclear, as all ultraviolet divergences are not influenced by 
temperature.

In the present paper we adopted the first point of view and dropped 
the zero-mode contribution, which is consistent also with earlier calculations
by Kutasov and Larsen \cite{kl}, as well as Klemm, Petkou and 
Siopsis \cite{klemm,siopsis}.
After this paper appeared in hep-th, this approach
was criticized by Dowker in
Ref.~\cite{dowker}, who argued in favor of the second convention, of keeping
the zero-mode contribution.
Dowker says that for the $d=2$ conformal scalar,
instead of Eq.~(\ref{ltemplimd2scalar}), we should have
 (there is an overall factor of 2 difference because he uses
parity-degenerate states)
\begin{equation}
U=-\frac1{12a}+\frac2a\sum_{n=0}^\infty{}'\frac{n}{e^{n\beta/a}-1},
\end{equation}
so there is an extra $T$  term at low temperature which of course cancels
the $-T$ term at high temperature. [See Eq.~(\ref{htemplimd2scalar}).]
 Of course his remark about the
Poisson sum formula being the basic mechanism does not establish
the existence of the zero-mode contribution; rather, it just relates
the two limits.  Let us try to determine which convention is preferable by 
thermodynamical considerations.

The first question is, where does the $n=0$ term come from?  Of course
there is a zero mode for a circle.  But its contribution to the partition
function seems ambiguous:
\begin{equation}
\ln Z_{\rm zero \, mode}=-\ln\left(1-e^{-\beta 0}\right)=-\ln\beta\cdot 0,
\label{zm0}
\end{equation}
so if one does not worry about the meaning of this, one gets
\begin{equation}
E_{\rm zero \, mode}=\frac{\partial}{\partial \beta}\ln\beta=\frac1\beta,
\label{zm}
\end{equation}
as Dowker states.  This does not seem very well-defined.
We would interpret Eq.~(\ref{zm0}) as an infinite constant,
which should be dealt with by the process of renormalization.  
Furthermore, in Sec.~III, we presented
an independent derivation of the energy starting from the Green's function
which directly gave the high-temperature limit (\ref{general2drep}) 
with the $-T$ there.
It seems clear there is a certain level of ambiguity here.

This ambiguity is really only a mathematical one.
Physically the situation is clear-cut: on a circle the eigenvalues are
$|n|/a$,  and thus zero for $n=0$. This means that  $n=0$  cannot contribute to
the sum at all; this case is eliminated physically from the start.

To completely resolve this ambiguity we note
 that such a term is in conflict with fundamental
thermodynamical principles.  If for low temperature
\begin{equation}
E=U-E_0=T=\frac\partial{\partial\beta}\beta F,
\end{equation}
the free energy must be, consistent with Eq.~(\ref{zm0}),
\begin{equation}
F=\frac1\beta\left(\ln\beta+ \mbox{constant}\right),
\label{dowkerfe}
\end{equation}
which then means that the entropy is
\begin{equation}
S=\beta(E-F)=1-\ln\beta- \mbox{constant},
\end{equation}
which diverges to $-\infty$
as $\beta\to\infty$. This grossly violates the Third Law
of thermodynamics. In contrast, our temperature correction
vanishes exponentially in that limit. [See Eq.~(\ref{lowtemp2dentropy}).]

Dowker bases his analysis on his earlier Ref.~\cite{dowker88}.  There
indeed he obtains the free energy in the form (\ref{dowkerfe}) as a
singular limit---i.e., the constant is infinite.  He further mentions the
entropy, saying that dropping the divergence is just a shift of the zero
of entropy and hence is unobservable, and then stating that ``it is not
clear whether or not there is a problem with Nernst's theorem.''  Evidently
there is such a problem, and appealing to the nonexistence of an ideal gas
state seems quite beside the point.

After this paper was completed, a preprint \cite{elizalde2} appeared which
completely supports our conclusions.

\end{document}